%% file: main.tex
\documentclass[conference]{IEEEtran}
\pdfoutput=1
\makeatletter
\def\ps@headings{%
\def\@oddhead{\mbox{}\scriptsize\rightmark \hfil \thepage}%
\def\@evenhead{\scriptsize\thepage \hfil\leftmark\mbox{}}%
\def\@oddfoot{}%
\def\@evenfoot{}}
\makeatother
\usepackage{cite,graphicx,epsfig, psfrag, subfigure, url, array}
\usepackage{algorithm, algorithmic, calc, color}
\usepackage{graphics}
\usepackage{amssymb, amsmath}
\usepackage{epstopdf}
\usepackage{multicol,balance}
\usepackage{bbm}

\newcommand {\enorm}[1]{\left|\left|#1\right|\right|_{2}}

\newtheorem{definition}{Definition}

\begin{document}

\title{
Predictive Blacklisting
\\as an Implicit Recommendation System}

\author{
{\Large Fabio Soldo, Anh Le, Athina Markopoulou}\\
University of California, Irvine\\
{\em \{fsoldo, anh.le, athina\}@uci.edu}
}
\maketitle

\input{abstract}
\input{introduction}
\input{related_work}

\input{data_analysis}
\input{formulation}

\input{model}
\input{performance}
\input{conclusion}

\bibliographystyle{IEEEtran}
\bibliography{main}

\end{document}

%% file: abstract.tex
\begin{abstract}
A widely used defense practice against malicious traffic on the Internet is through blacklists: lists of prolific attack sources are compiled and shared. The goal of blacklists is to predict and block future attack sources.
Existing blacklisting techniques have focused on the most prolific attack sources
and, more recently, on collaborative blacklisting. 
In this paper, we formulate the problem of forecasting attack sources (also referred to as ``predictive blacklisting'') based on shared attack logs as an implicit recommendation system. We compare the performance of existing approaches against the upper bound for prediction, and we demonstrate that there is much room for improvement. Inspired by the recent Netflix competition, we propose a multi-level prediction model that is adjusted and tuned specifically
for the attack forecasting problem. Our model captures and combines various factors, namely: attacker-victim history (using time-series) and attackers and/or victims interactions (using neighborhood models). We evaluate our combined method on one month of logs from {\tt Dshield.org} and demonstrate that it improves significantly the state-of-the-art.


\end{abstract} 

%% file: introduction.tex
\section{introduction}
\label{sec_introduction}

A widely used defense practice against malicious traffic on the Internet today is through blacklists: lists of the most prolific
attack sources are compiled, shared, and eventually blocked. Examples of computer and network blacklists include IP and DNS blacklists to help block unwanted web content, SPAM producers, and phishing sites.
 Sites such as {\tt DShield.org} \cite{Dshield}, process firewall and intrusion detection system (IDS) logs contributed by hundreds victim networks worldwide, and compile and publish blacklists of the most prolific attack sources reported in these logs.

Blacklists essentially attempt to forecast future malicious sources based on past logs.
It is desirable that they are {\em predictive}, {\em i.e.,} include many of the malicious sources that will appear
in the future and as few false positives as possible.
 It is also desirable that the blacklist size is short, especially when the blacklist is used online for checking every flow on the fly.
Predicting future malicious activity accurately and in a compact way
is a difficult problem. Given the wide use of blacklists on one hand, and the inherent complexity of the problem on the other hand, it is surprising how little has actually been done so far to systematically treat this problem.

The two most common techniques are GWOL and LWOL, according to the terminology of \cite{Zhang2008}.
{\em LWOL} stands for ``Local Worst Offender List'': security devices deployed on a specific site keep logs of malicious activity, and a blacklist of the most prolific attack sources, in terms of target IPs, is compiled. This local approach, however, fails to predict attack  sources that have never previously attacked this site; in this sense, a local blacklist protects the network reactively rather than proactively. Meanwhile, {\em GWOL} stands for ``Global Worst Offender List'' and refers to blacklists that include top attack sources that generate the highest number of attacks globally, as reported at universally reputable repositories, such as \cite{SANS, Dshield}. A problem with this approach is that the most prolific attack sources globally might be irrelevant to some victim networks that do not provide the corresponding vulnerable services.

Recently, Zhang {\em et al.} \cite{Zhang2008} proposed a collaborative blacklisting technique
called  ``highly predictive blacklisting''(or {\em HPB}).  They studied flow logs from {\tt Dshield.org}, defined the victim-to-victim similarity graph, and applied an algorithm resembling the Google's PageRank algorithm to identify the most relevant attackers for each victim. The HPB approach improved over LWOL and GWOL and is, to the best of our knowledge, the first methodological development  in this problem area in a long time.

Our work builds on and improves over \cite{Zhang2008}. Throughout the paper we use the terms {\em attack forecasting} and {\em predictive blacklisting} (in the terminology of \cite{Zhang2008}) interchangeably. We formulate the problem using a different methodological framework inspired by the emerging area of {\em recommendation systems} (RS) \cite{RSsurvey, netflix, amazon, googlenews}. Based on shared security logs, we study malicious behavior at the IP level, {\em i.e.,} considering the (attacker IP source, victim IP destination, time) tuple. We predict future malicious activity based on the past and we construct predictive blacklists specifically for each victim. We exploit both temporal (attack trends) and spatial (similarity of attackers and victims) features of malicious behavior. One family of temporal techniques predicts future attacks using the time series of the number of reported attacks. Another family of spatial techniques explores neighborhoods of victims as well as of joint attackers-victims.
We analyze 1-month of {\tt Dshield.org} data and evaluate different candidate techniques.
 We optimize each technique independently and then combine them together. We show that the combined method significantly improves the performance, {\em i.e.}, increases the predictiveness, or ``hit count'', of the blacklists over baseline approaches. Specifically, it improves up to 70\% the hit count of the HPB scheme with an average improvement over 57\%.
Last but not least, the formulation of the problem as an implicit recommendation system opens the possibility to apply powerful methodologies from machine learning to this problem.


The rest of this paper is organized as follows. Section \ref{sec_related_work} discusses related work. Section \ref{sec_data_analysis} gives a brief overview of some key features of the {\tt Dshield.org} dataset. Section IV  formulates the attack prediction problem in the recommendation systems framework; it also motivates this study by showing the gap between state-of-the-art approaches and the upper bound (achieved by an offline algorithm.) Section V presents the specific temporal and spatial methods we use for prediction. Section VI evaluates the individual methods and their combination over the {\tt Dshield.org} dataset; the combined method significantly outperforms the current state-of-the-art approach. Section VII concludes and discusses open issues and future work.

%% file: related_work.tex
\section{Our Work in Perspective}
\label{sec_related_work}

The two {\bf traditional approaches} to generate blacklists, LWOL and GWOL, according to the terminology of \cite{Zhang2008}, have already been outlined in the introduction. They both select the most prolific attackers based on past activity recorded in logs of a single victim site (in the case of LWOL) or of multiple victim sites (in the case of the  GWOL.) Both approaches have pros and cons. The local approach is essentially reactive but can be implemented by the operator of any network independently. The global approach uses more information that may or may not be relevant to particular victims, and requires sharing of logs among multiple victims, in a distributed way or through central repositories. There are also variations of these approaches, depending on whether a ``prolific'' attacker is defined based on the number of attacks launched (number of logs) or on the number of unique victims attacked.

Beyond the traditional approaches, the {\bf state-of-the-art} method today is the ``highly predictive blacklisting'' (HPB), recently proposed by Zhang {\em et al.} \cite{Zhang2008}. The main idea was that a victim should predict future attackers based not only on his own logs but also on logs of a few other ``similar'' victims. Similarity between two victims was defined as the number of their common attackers, based on empirical observations made earlier by Katti {\em et al.} \cite{Katti2003}. A graph that captures the similarity of victims was considered, and an algorithm resembling Google's PageRank was run on this graph to determine the relevance of attackers for a victim. In essence,  predictive blacklisting was posed as a link-analysis problem, and the focus was on relevance propagation on the victim-victim graph.

Compared to HPB \cite{Zhang2008}, our work solves the same problem (predictive blacklisting based on shared logs), but we have several important differences in methodology and intuition. We makes the following {\bf contributions}: (1) We formulate the problem as an {\em implicit recommendation system (RS)} \cite{RSsurvey} rather than as a link-analysis problem; this opens the possibility to apply a new set of powerful techniques from machine learning. Within the RS framework, we combine a number of specific techniques that capture and predict different behaviors present in our {\tt Dshield.org} dataset. Recall that our data are of the form (attacker IP address, victim IP address, time). (2)  One set of techniques are {\em spatial}, {\em i.e.,} use the notion of similarity of victims and/or attackers. HPB is a spatial case, where similarity is considered only among victims and is defined as the number of common attackers. (2a)  We use a different notion of victim-victim similarity
which focuses on simultaneous attacks from common sources (attacks performed by the same source at about the same time induce stronger similarity among victims.)
 (2b) Furthermore, we also define another notion of neighborhood that takes into account blocks of attackers and victims jointly, using 
 a co-clustering algorithm called cross-association (CA) \cite{Chakrabarti2004}. (3) Another set of techniques use time series to exploit {\em temporal trends} for prediction; to the best of our knowledge, this axis has not been exploited before for predictive blacklisting.
 This includes LWOL as a special case, where the past consists of a single time period.

Our proposal could be  {\bf implemented} at the shared logs repository, {\em i.e.}, it could be used, as an improvement of HPB, which is currently provided as a service by {\tt Dshield.org}; alternatively sharing and prediction could be implemented in a distributed way among collaborating participants. 
However, the goal  of this paper is the design and evaluation of the prediction algorithm  and not the development of a prototype.

Our work falls within the category of {\bf behavioral analysis}, in the sense that inferences are made based on flow logs as opposed to packet payload. However, we are interested in prediction and {\em not} in traffic classification \cite{blinc} or distinguishing legitimate from malicious traffic \cite{feamster-ccs,feamster-usenix},
{\em i.e.}, we work with flow logs that have already been classified as malicious by IDS and we focus on prediction.

Our evaluation is based on the {\tt Dshield.org} {\bf dataset}, which, despite its imperfections ({\em e.g.,} noise), has become a common reference used by many researchers in this area, including the state-of-the-art HPB \cite{Zhang2008} and other studies of malicious behavior on the Internet \cite{barford-miniinfocom, barford-pam, Katti2003}.

Finally, our problem formulation is inspired by {\bf recommendation systems} (RS) \cite{RSsurvey}, which currently find applications on e-commerce web sites, such as NetFlix \cite{netflix} and Amazon \cite{amazon}, as well as
 on other areas such as Google News \cite{googlenews}. 
   The problem of attack prediction is best modeled as an {\em implicit}  recommendation system, where ``ratings'' are inferred (not given explicitly) by observing malicious activity, and recommendations are provided to victims about what addresses to block in the future. Another complication is that malicious activities (``ratings'') vary  over time, which is currently an active research area in the RS community. 

%% file: data_analysis.tex
\section{The Dshield Dataset: Overview and Key Characteristics}
\label{sec_data_analysis}

\begin{figure*}[t!]
\centering
\subfigure[A sample of malicious activity]{
\includegraphics[scale=0.30, angle=0]{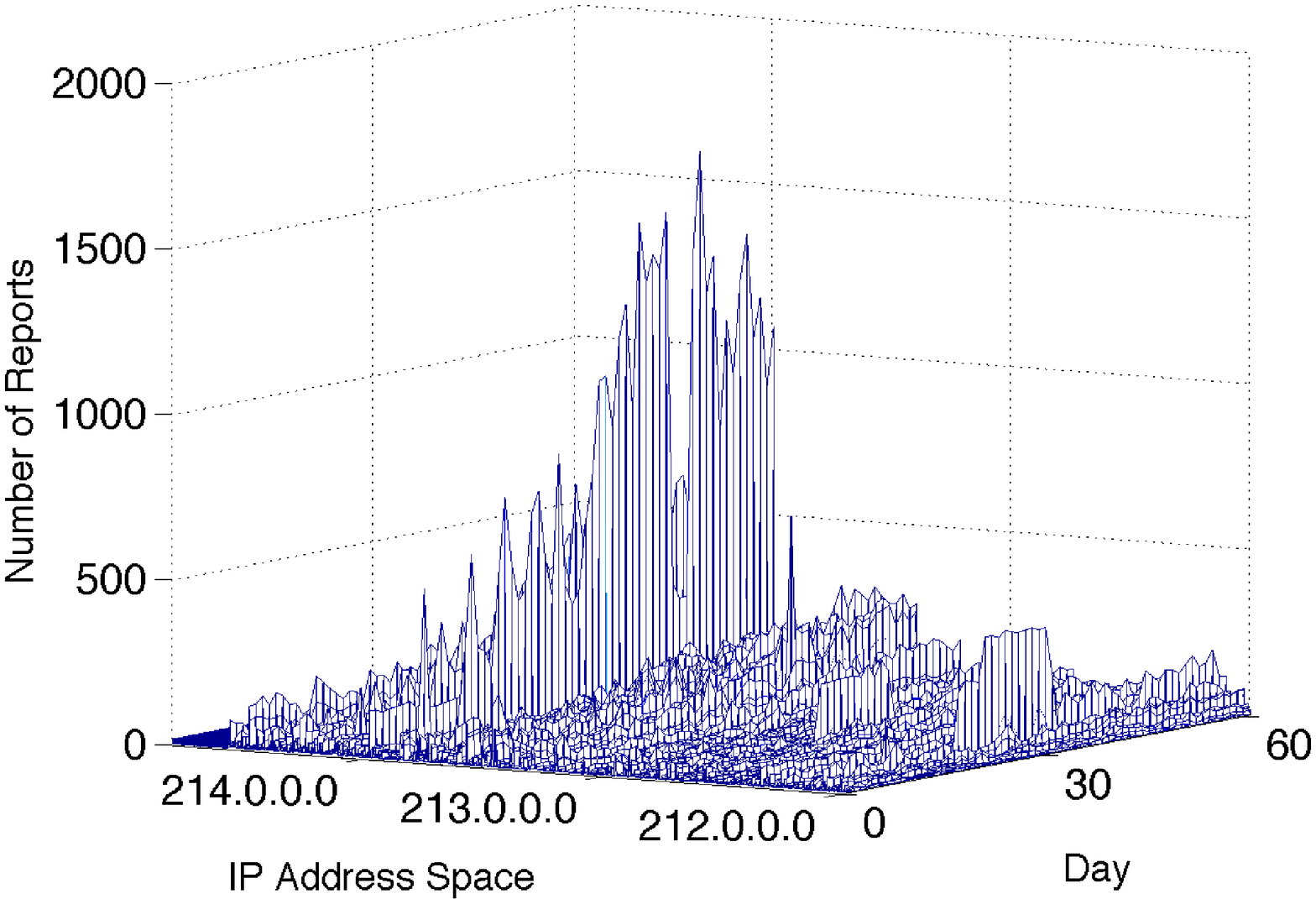}}
\subfigure[Temporal behavior: inter-arrival time of the same source appearing in the logs]{
\includegraphics[scale=0.33, angle=0]{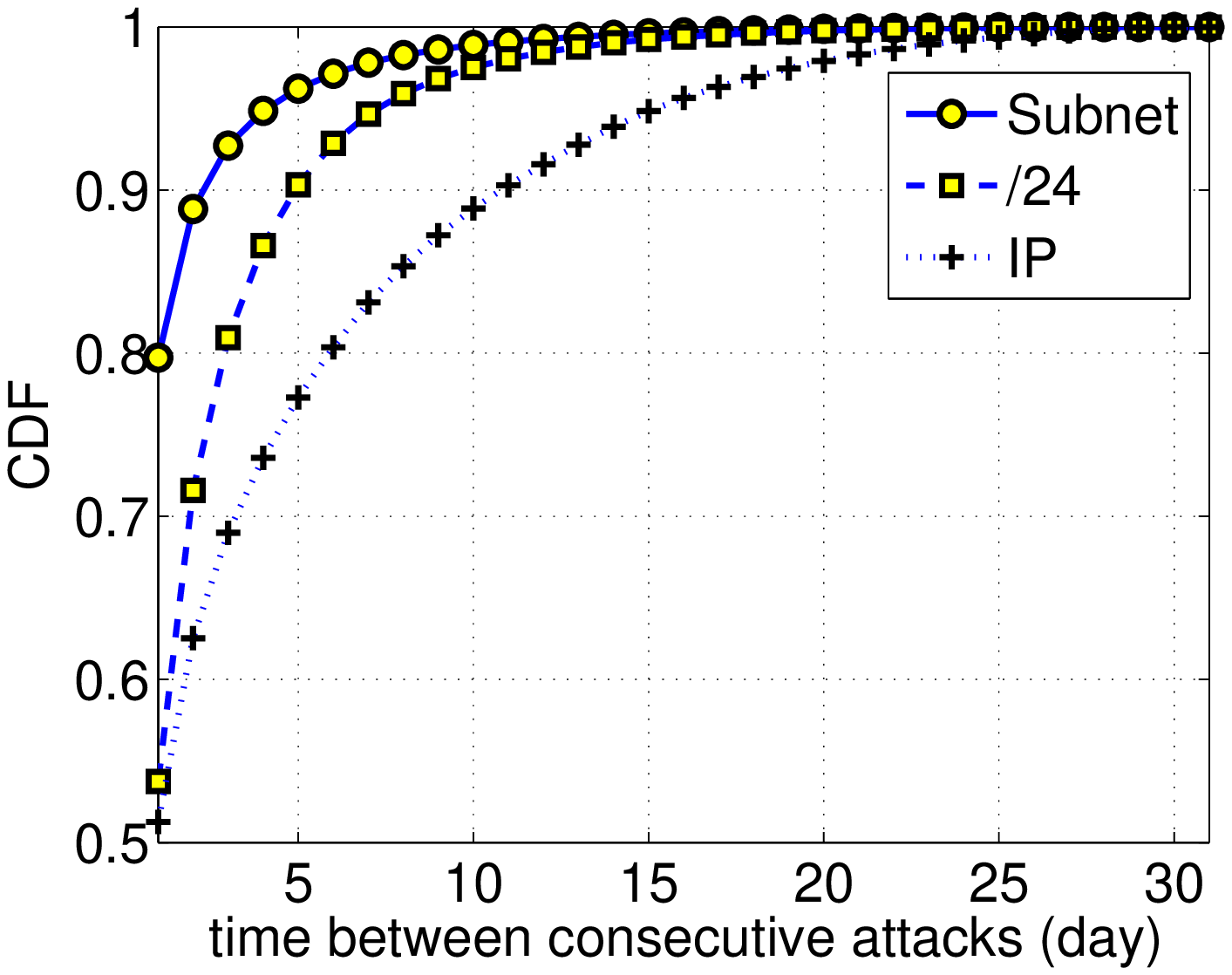}}
\subfigure[Common attackers among different victims]{
\includegraphics[scale=0.33, angle=0]{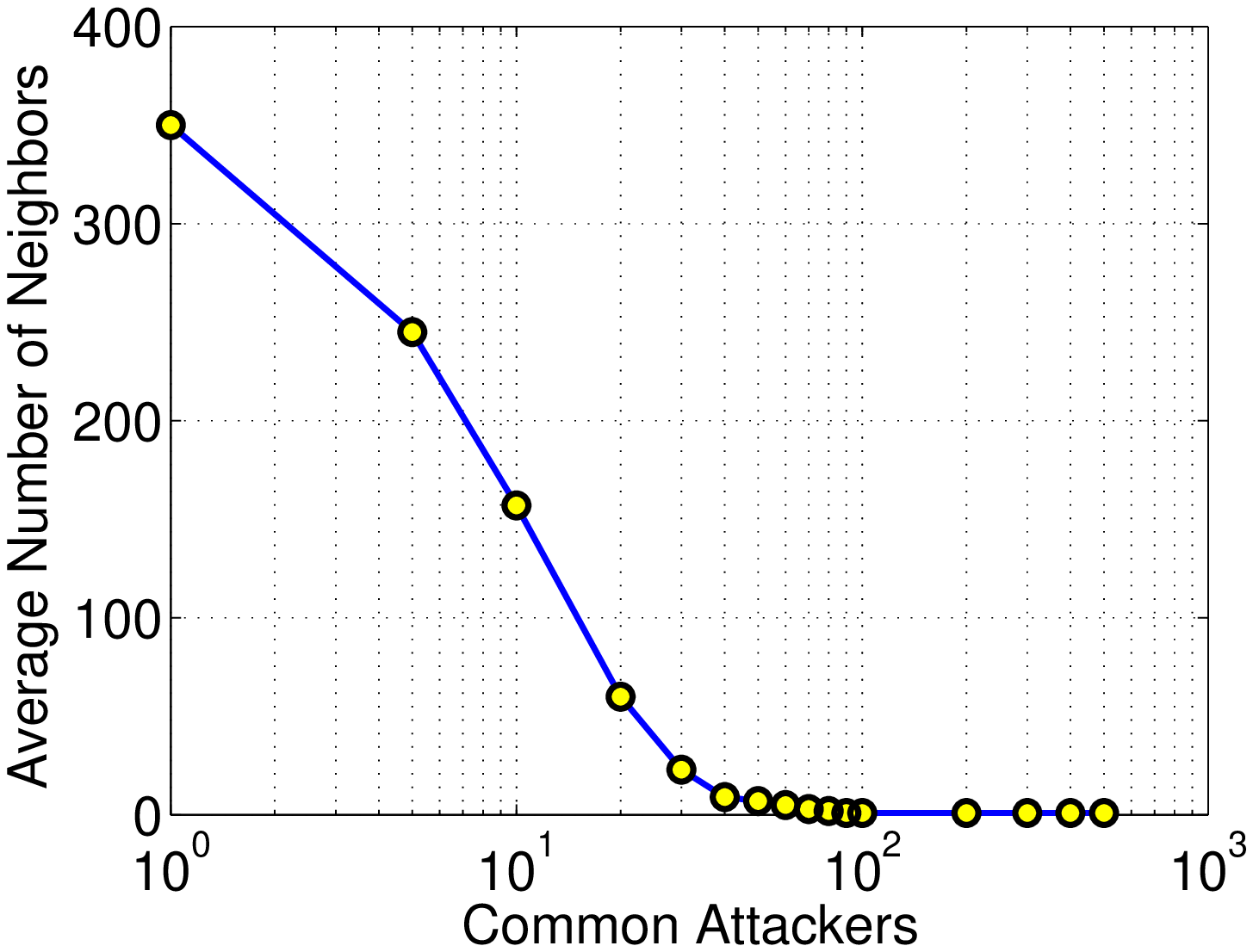}}
\caption{\label{fig:data3}Some insights from the Dshield dataset that motivated some of our design choices.}
\end{figure*}


In this study, we used logs from {\tt Dshield.org}\footnote{The authors are grateful to P.Barford and M.Blodgett from University of Wisconsin, Madison, for providing the dataset.} to understand the patterns existing in real data and to evaluate our prediction methods in practice. In this section, we briefly describe the dataset and mention some key properties that influenced the design of our prediction methods. We note, however, that data analysis is {\em not} the focus of this paper.\footnote{The thorough characterization of this dataset would be a measurements paper on its own, {\em e.g.,} see tech. report \cite{techreport} for details. Furthermore, the properties of {\tt Dshield} data have been studied by several researchers over the years, {\em e.g.,} see \cite{barford-pam, barford-miniinfocom, Katti2003}, and agree with our analysis as well. {\tt Dshield} data are used in this paper mainly to evaluate the methods developed here.}

{\bf The dataset.} {\bf Dshield} \cite{Dshield} is a repository of firewall and intrusion detection logs collected at hundreds of different networks all over Internet. 
The participating networks contribute their logs, which are then converted into a common format that includes the following fields: time stamp, contributor ID, source IP address, destination IP address, source port number, destination port number, and protocol number. In this paper, we work with the first three fields.
 One challenge when dealing with large-scale security log sharing systems is the amount of noise and errors in the data. For this reason, we pre-processed our data set to reduce noise and erroneous log entries, such as those belonging to invalid, non-routable, or unassigned IP addresses. Data from Dshield have been studied and used by several researchers over the years, such as \cite{barford-pam, barford-miniinfocom, Katti2003, Zhang2008, Soldo2009} to name a few examples, and the main findings ({\em e.g.} clustering of malicious sources and short-lived IP addresses in the logs) were consistent over time and with our own analysis \cite{techreport}.

 We analyzed 6 months of Dshield logs, from May to Oct. 2008  \cite{techreport}. In this paper, we present results for 1-month only (October 2008); the results were quite similar for the other months. The  pre-processed 1-month dataset consists of about 430M log entries, from $\sim$ 600 contributing networks, with more than 800K unique malicious IP sources every day.

{\bf Observations.} Fig. \ref{fig:data3} showcases some observations from the data that motivated design choices in our prediction. First, Fig. \ref{fig:data3}(a) offers a visualization 
of part of the data: it shows the number of logs generated by a portion of the IP space over time.
One can visually observe that there are several different activities taking place. 
Some sources attack consistently and by an order of magnitude higher than other sources (heavy hitters); some attack with moderate-high intensity but only for a few days (spikes); some attack continuously in a certain period  and do not appear again; finally, most other sources appear to be stealthy and generate limited activity.
The wide variety of dynamics in the same dataset poses a fundamental challenge for any prediction mechanism.
Methods, such as GWOL, focusing on heavy hitters will generally fail to detect stealthy activity. Methods focusing on continuous activity will not predict sudden spikes in activity. This motivated us to develop and {\em combine several complementary prediction techniques} that capture different behaviors.

Secondly, in Fig. \ref{fig:data3}(b), we show some information about the temporal behavior. In particular, we consider attack sources that appear at least twice in the logs and study the inter-arrival time between logs for the same attack source. We plot the cumulative distribution function (CDF)
of inter-arrival time  at three different levels of granularity: IP address, /24 prefix, and source subnet. We observe that for /24 prefixes, 90\% of attacks from the same source happen within a time window of 5 days while the remaining 10\% are widely spread over the entire month. Similar trends are true for the other levels.  
This implies that attacks have a short memory: if an attacker attacks more than once, then with high probability it will attack again soon. This motivated the {\em EWMA time series} approach we use for temporal prediction.

Another important aspect influencing the design of our prediction methods is the correlation among attacks seen by different victim networks. Let us call two victim networks ``neighbors'' if they share at least a certain number of common attackers. Fig. \ref{fig:data3}(c) shows the average number of neighbor networks as a function of this number of
common attacking IPs for a given day. Most victims share only a few attackers because there are a few source IPs (heavy hitters) that constantly attack most victim networks.
However, if we consider a strict definition of neighbors, {\em i.e.}, sharing a large number of attackers, each victim has a smaller number of neighbor, which is likely to capture a more meaningful type of interaction.
This motivated us to consider small neighborhoods ($\sim 25$ nodes) in our spatial prediction methods.





%% file: formulation.tex
\section{Problem Formulation and Framework}
\label{sec:formulation}

Our goal is to predict future malicious IP traffic based on past logs contributed by multiple victims.
 Predicting malicious IP traffic is an intrinsically difficult problem due to the variety of exploits and attacks taking place at the same time and the limited information available about them.

\subsection{Recommendation Systems vs. Attack Prediction}

In this paper, we frame the problem of attack prediction as an implicit recommendation system problem, as depicted in  Fig. \ref{fig:RS}. Recommendation systems (RS) aim at inferring unknown user rating about items from known (past) ratings. An example is the Netflix recommendation system, Netflix Cinematech, which aims at predicting unknown user ratings about movies from known ratings, in order to provide movie recommendations to its customers. Other examples of real deployment of RS include the Amazon recommendations \cite{amazon}, Google news personalization \cite{googlenews}, TiVo, GroupLens, to mention a few.  What makes the prediction possible is that ratings are not given randomly but according to a complex and user-specific rating model, which is not known in advance. The rating matrix is a result of several superimposed processes, some of which  are intuitive, while others need to be unveiled and confirmed through an accurate analysis of the dataset. For instance, a user can give on average a higher score to drama movies than to comedy, or vice versa, according to his preference; or he can rate differently on weekdays than on weekends, which might reflect his state of mind \cite{kdd09}.

We also aim at predicting future attacks leveraging  observed past activities.
Given a set of attackers and a set of victims, a number $r$ is associated with every (attack source, victim destination, time) triplet according to the logs: $r$ can indicate, for example, the number of time an attacker has been reported to attack a victim over a time period.
 More generally, we interpret $r$ as the rating or preference: higher $r$ indicates that an attacker intensifies its effort to attack a specific victim, thus implying a higher preference.  There are some important differences from a traditional RS. First, the intensity of an attack may vary over time, thus leading to a time-varying rating matrix. This poses a significant challenge to the direct application of traditional RS techniques that deal which static matrices. Secondly, the rating in this case is implicit, as it is inferred by activity reported in the logs, as opposed to ratings in RS explicitly provided by the users themselves.


In the rest of this section, we first formalize the analogy between  recommendation systems and attack prediction.
Then, we define upper bounds for prediction and quantify the gap that exists today between the state-of-the-art prediction and what is actually achievable.
In subsequent sections, we propose specific methods that bridge this gap.

\subsection{The Recommendation System Problem}

\subsubsection{Notation}

Let $\mathcal V$ be the set of users (customers) and $\mathcal A$  be the set of items.
In practical applications, these two sets  are usually very large, including up to hundreds  of thousands of elements.
A user is allowed to rate items to indicate how much she likes specific items.
Let $\mathcal R$ be the set of possible {\em ratings},  $\mathcal R = \{1,2,...,N\}$, where $N$ is typically a small integer.
Let $r_{u,i}$ be the rating assigned by user $u$ to item $i$ and $R$ be the entire $|\mathcal V|$-by-$|\mathcal A |$ rating matrix.

\begin{figure}[t!]
\begin{center}
\includegraphics[width=75mm]{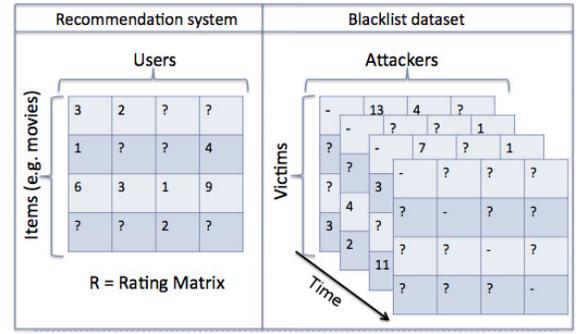}
\caption{\label{fig:RS} Analogy between Recommendation Systems (left) and Attack Prediction (right). The former infers unknown user ratings about items from known ones. The latter deals with time varying ratings.}
\end{center}
\end{figure}


\subsubsection{Problem Formulation}

A recommendation system (RS) aims at inferring unknown ratings from known ones, as shown in Fig. \ref{fig:RS}-left.
Ultimately, the goal of RS is to find for every user $u$, the item, $i_u$, that has the highest rating \cite{RSsurvey}:
$ i_u = \text{arg}\max_{i' \in \mathcal I} r_{ui'} \,, \forall u \in \mathcal V$.
What makes the RS problem difficult is that the values $r_{u,i}$'s are unknown.
We only know a limited subset of ratings because a user generally does not rate all available items but only a subset of them, typically orders of magnitude smaller than the number available items.

Let $\mathcal K_u$ be the set of items for which the rating $r_{u,i}$'s are known, and $\mathcal{\bar{K}}_u$ be its complement.
The RS problem can be formalized as follows:
\begin{equation}
\label{eq:RS}
\text{find} \quad i_u = \text{arg}\max_{i' \in \mathcal{\bar{K}}_u} r_{ui'} \qquad \forall u \in \mathcal V\,.
\end{equation}
 
 The recommended item, $i_u$, for user $u$,  maximizes Eq. (\ref{eq:RS}) and may be different for every user. The solution of  Eq. (\ref{eq:RS}) is usually obtained by first estimating the matrix $R$ on the subset $\mathcal{\bar{K}}_u$, and then, for every user, selecting  the item for which the {\em estimated}
 rating is the highest. In general, if we want to recommend $N\geq 1$ items we need to select the top-$N$ items for which the estimated ratings are the highest.




\subsection{The Attack Prediction Problem}

\subsubsection{Notation}

We denote with $ \mathcal V$ the set of victim networks and with  $\mathcal A$ the set of attackers ({\em i.e.}, source IP prefixes where attacks are launched from.)
Let $t$ indicate the time an attack was performed according to its log. Unless otherwise specified, $t$ will indicate the day the attack was reported. $T$ denotes the time window under consideration, 
so $t=1,2,...,T$. Moreover, we partition $T$ in two windows of consecutive days: a {\em training window},  $T_{train}$, and a {\em testing window} $T_{test}$,
to separate training data, used  to tune the prediction algorithm, $t \in T_{train}$, from testing data, used to validate the predictions, $t \in T_{test}$.

Similar to the RS problem, we define a  3-dimensional rating matrix $R$ so that per every tuple $(a,v,t)$, $r_{a,v}(t)$ represents the number of attacks reported on day $t$ from $a\in \mathcal A$ to $v\in \mathcal V$. 
 We denote with $B$ the binary matrix that indicates whether or not an attack occurred: $b_{a,v}(t) = 1$  iff $r_{a,v}(t) > 0$, and $b_{a,v}(t) = 0$ otherwise.
Finally, we indicate with $\mathcal A_v(T)$, the set of attackers that were reported by victim $v$ during the time period $T$: $$\mathcal A_v(T)= \{a\in \mathcal A \ : \  \exists t\in T \ \text{s.t.} \ b_{a,v}(t)=1\}$$
and with $\mathcal A(T)$ the total set of attack sources reported in $T$: $$\mathcal A(T) = \cup_{v\in \mathcal V} \mathcal A_v(T)$$




\subsubsection{Problem Formulation}

For every victim, $v$, we are interested in determining which attackers are more likely to attack $v$ in the future given  past 
observation of malicious activity.  In practice, this translates into providing a blacklist ($\mathcal{BL}$) of sources that are likely to attack in the future.  
Given a fixed blacklist size, $N$, let $\mathcal{\tilde{BL}}$ be any set of $N$ different attackers. The problem of attack prediction can be formalized as follows:
\begin{equation}
\label{AP}
\text{find} \quad \mathcal{BL}(v)  = \text{arg}\max_{\mathcal{\tilde{BL} \subset \mathcal A}}   \sum_{t\in T_{test}} \sum_{a \in \mathcal{\tilde{BL}}} b_{a,v}(t) 
\end{equation}


The output of the attack prediction problem is a set of blacklists,  $\mathcal{BL}(v) = \{a_{1}^v,a_{2}^v,...,a_{N}^v\} \subset \mathcal A$ customized for every victim, $v$, such that each blacklist, $\mathcal{BL}(v)$, contains the top $N$ attackers that are 
 more likely to attack $v$ in the time window $T_{test}$. 
 The difficulty of this problem is that for every $t \in T_{test}$, we need an entire $|\mathcal A|$-by-$|\mathcal V|$ matrix to be estimated before the $\max$ operation can be performed, as illustrated in Fig. \ref{fig:RS}-right. In this sense, this problem is a generalization of the recommendation problem,  Eq. (\ref{eq:RS}), where $R$ is now defined on 3-dimensional space, $  \mathcal V \times \mathcal A \times T$, rather than a 2-dimensional space. While the RS problem traditionally estimates missing elements  in a matrix, the attack prediction problem estimates matrixes  in a tensor.

Finally, we observe that per every blacklist $\mathcal{BL}$, and testing period, $T_{test}$, the total number of false positive ($FP$) can be defined as: 
 
 $$FP_{\mathcal{BL}}(T_{test}) =  \sum_{t\in T_{test}}\big(N - \sum_{a \in \mathcal{\tilde{BL}}} b_{a,v}(t) \big)$$
 
Thus, for fixed blacklist length $N$, solving Problem (\ref{AP}) is equivalent to finding the blacklist that minimizes the number of false positive. 

\subsection{\label{sec:motivation}Upper Bounds and State-of-the-Art}

Given a blacklist of length $N$, a metric of its predictiveness is the hit count, as defined in \cite{Zhang2008}: the  number of attackers in the blacklist that are correctly predicted, {\em i.e.}, malicious activity from these sources appears in the logs in the next time slot. A blacklist with higher hit count is more ``predictive.''


A future attacker {\em can} be predicted if it already appeared at least once in the logs of some victim networks. Clearly, we cannot accurately
predict attackers that have never been reported before.
Consequently, we can define two upper bounds on the hit count, a global and a local upper bound, depending on the sets of logs we use to make our prediction.

\begin{definition} [{\bf Global Upper Bound}] Using notations defined above,
 for every victim $v$, we define the global upper bound on the hit count of $v$, $GUB(v)$,
as the  number of attackers that are both in the training window of {\em any} victim and in the testing window of $v$:
\begin{equation}
GUB(v) =   \mathcal A(T_{train}) \cap  \mathcal A_v(T_{test})\,.
\end{equation}
\end{definition}
This represents the maximum number of attackers of $v$ that are predictable in $T_{test}$, given observations obtained in $T_{train}$.
This upper bound corresponds to the case that the past logs of {\em all} victims  are available to make prediction, as it is the case when using central repositories, such as  {\tt Dshield.org}, or when each victim shares information with all other victims.

\begin{definition} [{\bf Local Upper Bound}]
For every victim $v$, we define the local upper bound on the hit count of $v$, $LUB(v)$,
as the  number of attackers that are both in the training window and in the testing window of $v$:
\begin{equation}
LUB(v) =   \mathcal A_v(T_{train}) \cap  \mathcal A_v(T_{test})\,.
\end{equation}
\end{definition}
$LUB(v)$ represents the upper bound on the hit count when each victim, $v$,  only has access to its local security logs, but not to the logs of other victims. This is a very typical case in practice today.
Because $A_v(T_{train}) \subseteq A(T_{train})$, the following inequality holds trivially: $LUB(v)  \leq GUB(v)$. At the end of this section, we will quantify this gap on a real dataset of malicious sources.

\begin{figure}[t!]
\begin{center}
\includegraphics[width=75mm]{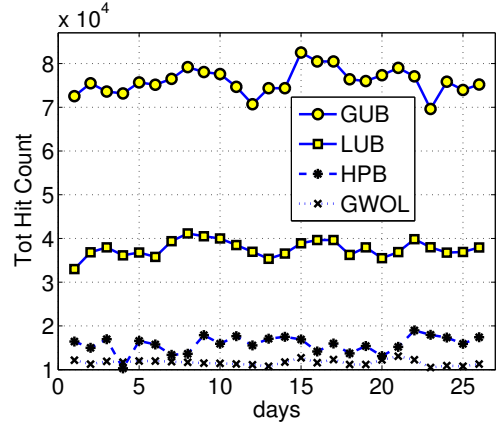}
\caption{Comparing different prediction strategies (in terms of total hit-count) on 1-month of {\tt Dshield.org} logs. Observe that the state-of-the art HPB improves over baseline, GWOL, but there is still a large performance gap until the upper bounds, LUB and GUB. }
\label{fig:motivation}
\end{center}
\end{figure}

{\bf State-of-the-art.} The next natural question is how far are state-of-the-art methods today from these upper bounds? Existing approaches for creating predictive blacklists include the traditional LWOL and GWOL, as well as the recently proposed state-of-the-art HPB \cite{Zhang2008}. These methods have been described in detail in the previous section \ref{sec_related_work}.

{\bf Room for improvement.} In Fig. \ref{fig:motivation}, we compare the total hit count (for all victims in the system) of different prediction strategies on 1-month of {\tt Dshield.org} logs. For a fair comparison, we require all methods to use the same blacklist length $N$.
To be consistent with \cite{Zhang2008}, we set $N=1,000$ and consider  that every source in the predictive blacklist is an IP prefix /24. We make two main observations.

First,  the state-of-the-art HPB strategy 
 brings benefit over the GWOL strategy. In our dataset, we observed an average improvement in the hit count  of about 36\% over GWOL, which confirms prior results on older data \cite{Zhang2008}. However, the gap between HPB and both LUB and GUB is still significant! This shows that there is a large room for improvement in attack prediction, which remains unexplored. This gap motivated us to further investigate this problem in this paper.

The second observation is the large gap between LUB and GUB.  This quantifies the improvement in attack prediction when different victim networks share their logs on observed malicious traffic.
Collaboration between different networks becomes a crucial factor when dealing with attack prediction because 
more shared information can potentially reveal correlation between attacks that cannot be discovered otherwise.

%% file: model.tex
\section{Model Overview}

Motivated by the observations made in Sections \ref{sec_data_analysis} and \ref{sec:formulation}, we develop a multi-level prediction
framework to capture the different behaviors and structures observed in the data.

\subsection{Time Series for Attack Prediction}		

A fundamental difference between forecasting attacks and typical recommendation systems is the way the temporal dynamics affect the ratings. In recommendation systems, rating are given at different times, 
but  once given they cannot change.
The goal is then to use the known ratings as ground truth and estimate the missing ratings based on them.
In contrast, in the  attack prediction problem, ``ratings'' vary rapidly over time as they represent the number of attacks (logs) reported in different days.
As a consequence, in order to be able to forecast attacks, we must account not only for the time an attack was reported but also for its evolution over time.

Every rating $r_{a,v}(t)$ is essentially a signal over time. We use a time series approach to model the temporal dynamics associated with every pair $(a,v)$.
As observed in the data (Fig. \ref{fig:data3}(b)),  multiple attacks from the same source happen within a small time interval from each other, {\em i.e.}, for the large majority of attacking IP prefixes, the future activity strongly depends on the recent past.
Motivated by this observation, we use an Exponential Weighted Moving Average (EWMA) model, which predicts future values based on past values weighted with exponentially decreasing weights toward older values.

We indicate with $\tilde r^{TS}_{a,v}(t+1)$ the predicted value of $ r_{a,v}(t+1)$  given the past observations, $ r_{a,v}(t')$, at time $t'\leq t$.
We  estimate $\tilde r^{TS}_{a,v}(t+1)$ as
\begin{equation}
\label{ewma} \tilde r^{TS}_{a,v}(t+1) = \sum_{t'=1}^{t}\alpha( 1-\alpha )^{t-t'} r_{a,v}(t')\,,
\end{equation}
where $\alpha\in (0,1)$ is the smoothing coefficient, and  $t'=1,...,t$ indicates the training window, where 1 corresponds to the oldest day considered, and $t$ is the most recent one.

We note that Eq. (\ref{ewma})  include as a special case the LWOL when the training window has only one day, $t = 1$.
   In this case, the EWMA model gives the highest weight to attackers that were most prolific in the last day, as in the LWOL strategy.
However, the general EWMA model has higher flexibility that can model temporal trends.
Weights assigned to past observations are exponentially scaled so that older observations have smaller weights.
This allows to account for spikes in the number of reports, which are frequently observed in our analysis of malicious traffic activities.

We also observed that an improved prediction accuracy can be obtained when applying the same EWMA model to the binary version of $R$, $B$.
This is because attackers that performed a large number of attacks in the recent past are not the most likely to also attack in the future.
An attacker can stop its activities at any time. Moreover,  the number of reports might also be very sensitive to the specific configuration of the victim NIDS.
The group of attackers that is more likely to keep on attacking is the one that was continuously reported as malicious for a large number of days  independently from the number of reports.
Therefore, an improved forecast can be obtained based on $B$:
\begin{equation}
\label{ewmab} r^{TS}_{a,v}(t+1) = \sum_{t'=1}^{t}\alpha( 1-\alpha )^{t-t'} b_{a,v}(t')\,,
\end{equation}
where $ r^{TS}_{a,v}(t+1)$ indicates the forecast for $ b_{a,v}(t+1)$ and can be interpreted as a measure of how likely an attacker is to attack again given its past history. In the rest of this paper, we will use the improved forecast based on $B$ when we mention the Time Series (TS) method.

\subsection{Neighborhood Model}

The strategies described above can model simple temporal dynamics accurately and with low complexity. However, a prediction solely based on time will fail to capture spatial correlations between different attackers and different victims in the IP space. {\em E.g.},
 a persistent attacker that switches its target every day may easily evade this level of prediction. In this section, we show how to capture such ``spatial'' patterns and use them for prediction. We define two types of neighborhoods: one that captures the similarity of victims ($k$NN) and another that captures joint attacker-victim similarity (CA).

\subsubsection{Victim Neighborhood ($k$NN)}
One of the most popular approaches in recommendation systems is the use of neighborhood models.
Neighborhood models build on the idea that predictions can be made by trusting similar peers.
For instance, in a movie recommendation system, a neighborhood model based on user similarity will predict that user John Smith likes
Harry Potter, only if users that have shown {\em similar} taste to John Smith and have already seen Harry Potter,
liked it. 

In this context,  the definition of similarity plays a fundamental role. There are several different similarity measures proposed in the literature.
The most commonly used is the Pearson correlation, which generalizes the notion of cosine distance of vectors with non-zero mean. Formally, given two $n$-dimensional vectors, $x,y$, with mean values, $m_x$, $m_y$, respectively,  the Pearson correlation of $x$ and $y$ is defined as
\begin{equation}
s_{xy} = \frac{ \sum_{i=1}^n  (x_i - m_x)(y_i - m_y) } {\sqrt{\sum_{i=1}^n (x_i - m_x)^2} \sqrt{\sum_{i=1}^n (y_i - m_y)^2 }}\,.
\end{equation}
We observe that for zero mean vectors, this reduces to $s_{xy} =\frac{ x\cdot y}{\enorm x \enorm y} = cos(x,y)$.

In this work, we developed a variation of the Pearson similarity to account  for the time the attacks were performed.
This is also motivated by \cite{Katti2003}, which observed that victim networks, that persistently share common attackers, are often attacked at about the same time.
 For every pair of victims, $u,v$ we define their similarity, $s_{uv}$, as
\begin{equation}
\label{s} s_{uv} = \sum_{ t_1 \leq t_2\in T_{train}} e^{-|t_2-t_1|} \frac{  \sum_{ a\in \mathcal A}b_{a,u}(t_1)\cdot b_{a,v}(t_2) }{   \enorm{ b_u(t_1)  } \enorm{ b_v(t_2)  } }\,,
\end{equation}
where 
$\enorm{ b_u(t_1)  } = \sqrt{ \sum_{a\in \mathcal A}  b^2_{a,u}(t_1)}$. Notice that if $u$ and $v$ report attacks at the same time, $s_{uv}$ reduces to a sum of cosine similarities. When $u$ and $v$ report attacks by the same attacker at different times, the smoothing factor, $e^{-|t_2-t_1|}$, accounts for the time interval between the two attacks.

We tried several similarity measures, and we found that the one in Eq. (\ref{s}) worked best.  Attacker activities might vary broadly over time. Eq. (\ref{s}) models the intuition that victims, that share attacks from the same source in the same time slot, are more similar to each other than victims sharing common attackers but during very different time since they are more likely affected by the same type of attack. Thus, giving higher importance to attacks occurring in the same time slot captures a stronger correlation among victims than just using the number of common attackers.

We adapt a $k$-nearest neighbors ($k$NN) model to the attack prediction problem.
The idea of traditional $k$NN model is to model missing ratings as a weighted average of known rating given to the {\em same item} by similar users:
\begin{equation}
	\label{knn} r^{kNN}_{a,v}(t) = \frac{\sum_{ u\in N^k(v;a)} s_{uv} r_{a,u}(t)}{ \sum_{u\in N^k(v;a)} s_{uv}  },\quad \forall t \in T_{test}
\end{equation}
where, $r^{kNN}_{a,v}(t)$ is the prediction provided by the $k$NN model, and $N^k(v;a)$ represents the neighborhood of top $k$ similar victims to $v$ according to the similarity measure, $s$, for which $r_{a,u}(t)$ is known.

In order to compute $ r^{kNN}_{a,v}(t) $, we need two main ingredients: a similarity measure between victims, $s$, and a set of known rating for the attacker $a$, $r_{a,u}(t)$.
What prevents us from a direct application of Eq. (\ref{knn}) is that none of the ratings, $r_{a,u}(t)$, is known in the testing window. Thus, the neighborhood $N^k(v;a)$ is empty.
To overcome this difficulty, we leverage
the forecast provided by the time series approach in Eq. (\ref{ewmab}): 
 \begin{equation}
	\label{knnTS} r^{kNN}_{a,v}(t) = \frac{\sum_{ u\in N^k(v;a)} s_{uv} r^{TS}_{a,u}(t)}{ \sum_{u\in N^k(v;a)} s_{uv}  },\quad \forall t \in T_{test}\,,
\end{equation}
which 
is a generalization of the $k$NN model.

\subsubsection{Joint Attacker-Victim Neighborhood (CA)}

\begin{figure}[t!]
\begin{center}
\includegraphics[width=65mm]{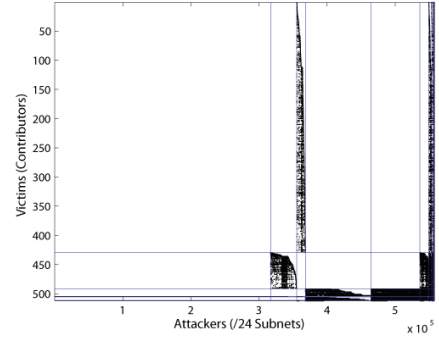}
\caption{Result of applying the CA algorithm on 1-day Dshield logs. A rectangular block indicates a group of similar sources and victims identified by the CA.}
\label{fig:resultCA}
\end{center}
\end{figure}

In addition to the victim neighborhood explored by the $k$NN model, we also studied the joint neighborhood of attackers and victims. Our intuition is that not only victim similarity but also the similarity among the attackers should be considered when constructing the blacklists.
For example, consider botnets, which are the main source of malicious activity on the Internet today: machines in a botnet typically attack the same set of victims. However, the timing of the attacks might differ due to different phases of the attacks \cite{Rajab2006, Cooke2005}: typically a scanning phase is carried out by a few machines before the attacking phase, which might be carried out by more machines, {\em e.g.}, in an instance of distributed denial-of- service (DDoS) attack. Therefore, knowing the similarity among the machines of a botnet, even if only a few of them are detected by a victim's IDS, enables the victim to preemptively put the other ``similar'' machines of the botnet into his blacklist.

To find similarity among both victims and attackers simultaneously, we apply the cross-associations (CA) algorithm \cite{Chakrabarti2004} -- a fully automatic clustering algorithm that finds row and column groups of sparse binary matrices. In this way, we find groups of both similar victims (contributors) and attackers (/24 subnets.) Fig. \ref{fig:resultCA} depicts the result of applying the CA on a contributor-subnet matrix of 1-day log data. (The original binary matrix describing the attacker-victim activity is omitted due to lack of space. For more information about the use of the CA algorithm for analyzing {\tt Dshield} logs, we refer the reader to our technical report \cite{Le2009}.) On average, the CA finds over 100 groups per day.

For each group (depicted as a rectangular block in Fig. \ref{fig:resultCA}), we calculate its density as the ratio between the occupied area and the total area of the rectangle. 
Then, we use the density of a group to quantify the strength of correlation among the attackers and victims within the group. Intuitively, a dense group corresponds to an attacker-victim bipartite graph that resembles a complete bipartite graph, thus indicating strong correlation. Finally, we use this density for forecast: the denser a group is, the more likely its attackers will attack its victims.

More formally, $\tilde r^{CA}_{a,v}(t+1) = \rho_{a,v}(t)$, where $\rho_{a,v}(t)\in [0,1]$ is the density of the group that contains the pair $(a,v)$ at time $t$. We can further  improve this CA-based prediction by capturing the persistence of the attacker and victim correlation over time. In particular, we apply the EWMA model on the time series of the density to predict the rating. The intuition is that if an attacker shows up in a neighborhood of a victim persistently, he is more likely to attack again the victim  than other attackers. Formally,
\begin{equation}
 r^{CA}_{a,v}(t+1) = \sum^t_{t'=1} \alpha (1 - \alpha)^{t-t'}\rho_{a,v}(t')\,.
\end{equation}
Our empirical study shows that the EWMA-CA prediction can improve the hit count by 25\% over the simple CA prediction. 








\subsection{ Combine Predictors}

The combination of different predictors (either obtained from different methods, or with the same method trained on different subset of the data) is generally referred to as ensemble learning.
The idea of ensemble learning is rooted in the traditional wisdom that ``in a multitude of counselors there is safety'' \cite{elder96}. 
Although the gain of ensemble learning is not fully understood yet, it is generally acknowledged  that such an approach is particularly suited in scenarios where a complex
system is better explained by the combination of different phenomena, which results in different structures in the data, rather than by a single phenomenon, {\em e.g.}, see Ch. 13 and 18 of \cite{ensemble}.
 The diverse dynamics observed in the analysis of malicious traffic motivated us to combine diverse algorithms, such as the time series approach to model temporal trends, the $k$NN to model victims similarity, and the CA clustering algorithm to model persistent groups of attackers-victims.

There are different methods to combine predictors. A typical approach is to consider the average of individual predictors.
What we found more effective is to (i) use the time series prediction as a base predictor and (ii) weight the neighborhood models with weights proportional to their accuracy. More specifically, for $k$NN we define
$$w_{a,v}^{kNN} = \frac{  \sum_{u\in N(v;a)} s_{uv} }{ \sum_{u\in N(v;a)} s_{uv} + \lambda_1}$$ where $\lambda_1$ is a parameter that needs to be estimated. 
 The intuition is that we want to rely more on $k$NN when $v$ has a strong neighborhood of similar victims.
 When $\sum_{u\in N(v;a)}s_{uv}$  is small, {\em i.e.,} only a neighborhood of poorly similar victims is available,  we prefer instead to rely on other predictors.
Similarly, we define a weight for the CA algorithm, $$w_{a,v}^{CA} = \frac{ \sum_{t\in T_{train}}\rho_{a,v}(t)}{  \sum_{t\in T_{train}}\rho_{a,v}(t)+ \lambda_2 }$$
so that, $w_{a,v}^{CA}\simeq 1$ for a pair $(a,v)$ that belongs to dense clusters; $w_{a,v}^{CA}\simeq 0$ when the density is low.

In summary, our rule for combining all methods together and giving a single rating/prediction is the following:

\begin{equation}
\label{combination} \hat b_{a,v}(t)  = r^{TS}_{a,v}(t)  + w_{a,v}^{kNN} r^{kNN}_{a,v}(t)  +w_{a,v}^{CA} r^{CA}_{a,v}(t)
\end{equation}

where  $\hat b_{a,v}(t)$ is the estimated value of $ b_{a,v}(t), \ \forall t\in T_{test}$.

%% file: performance.tex
\section{Performance Evaluation}


\subsection{Setup}

{\bf Data set.} We evaluate the performance of our prediction algorithm using 1-month of real logs on malicious IP sources provided by {\texttt Dshield.org}, as described in detail in Section \ref{sec_data_analysis}.

{\bf Metrics.}
We use two different metrics to evaluate the predictiveness of the blacklisting methods: the total hit count and the prediction rate.
The {\em hit count} was defined in Section \ref{sec:motivation} and represents the number of attackers in the blacklist that are correctly predicted; it is bounded by the blacklist length itself. When the algorithm provides individual victims with their customized blacklist, the total hit count is defined as the sum of the hitcounts over all contributors. The {\em prediction rate} is the ratio between the hit count and the global upper bound for each contributor separately. Thus, for each contributor, the prediction rate is a number in $[0,1]$, which represents the fraction of attackers correctly predicted out of the global upper bound of the contributor, which is the maximum number of attackers that can be predicted based on past logs. A prediction rate 1 indicates perfect prediction.

{\bf Parameters.} We call the time period (in the recent past) over which logs are analyzed in order to produce the blacklist {\em the training window}. We call the time period (in the near future) for which the forecast is valid
{\em the testing window}. Unless otherwise specified,  we use a  5-day training window and 1-day testing window. We motivate these choices in Section \ref{sec:windows}. Parameters $\alpha$, $\lambda_1$ and $\lambda_2$ are estimated using leave-one-out cross validation on the training set. 
Finally, for a fair comparison with prior work \cite{Zhang2008}, each predictive blacklist specifies /24 IP prefixes, which is also often the case in practice \cite{Dshield}. However, we note that the methodology described here applies to {\em any granularity} of IP address/prefix considered.


{\bf Complexity.} The complexity of the combined prediction depends on the complexity of the individual methods. Computing the TS prediction requires $O(T_{train})$ operations for each rating $r_{a,v}^{TS}$. Thus, its overall complexity is $O(T_{train}|\mathcal A||\mathcal V|)$.  The complexity of the $k$NN model is the computation of the similarity matrix $O(T_{train}|\mathcal V||\mathcal V|)$ plus the complexity of computing Eq.(\ref{knn}) for every pair $(a,v)$, that is $O(k|\mathcal A||\mathcal V|) = O(|\mathcal A||\mathcal V|)$ since $k$ is a constant. 
Finally, the CA clustering is a heuristic algorithm with a complexity empirically observed to be bounded by $O(|\mathcal A||\mathcal V|)$ \cite{Chakrabarti2004}. In practice,  $|\mathcal V|$ is  orders of magnitude smaller than $|\mathcal A|$, thus the overall asymptotical  complexity is bounded by $O(T_{train}|\mathcal A||\mathcal V|)$, that is, it increases linearly with the size of the data set, $R$. In our experiments, we could compute predictive blacklists for all contributors in $\sim$20 minutes with a 2.8 GHz processor and 32 GBs of RAM.

\subsection{Performance Evaluation and Comparison of Methods}

We group prediction schemes in two categories depending on whether they use local or global information. In the local category, there are the time series (TS) and LWOL, since they both use only local logs available at each network. In the global category belong the neighborhood models, such as $k$NN and EWMA-CA, as well as GWOL, since they use logs collected at shared among multiple networks.

 \begin{figure*}[t!]
\centering
\subfigure[Local approaches: TS and LWOL]{
\includegraphics[width=58mm]{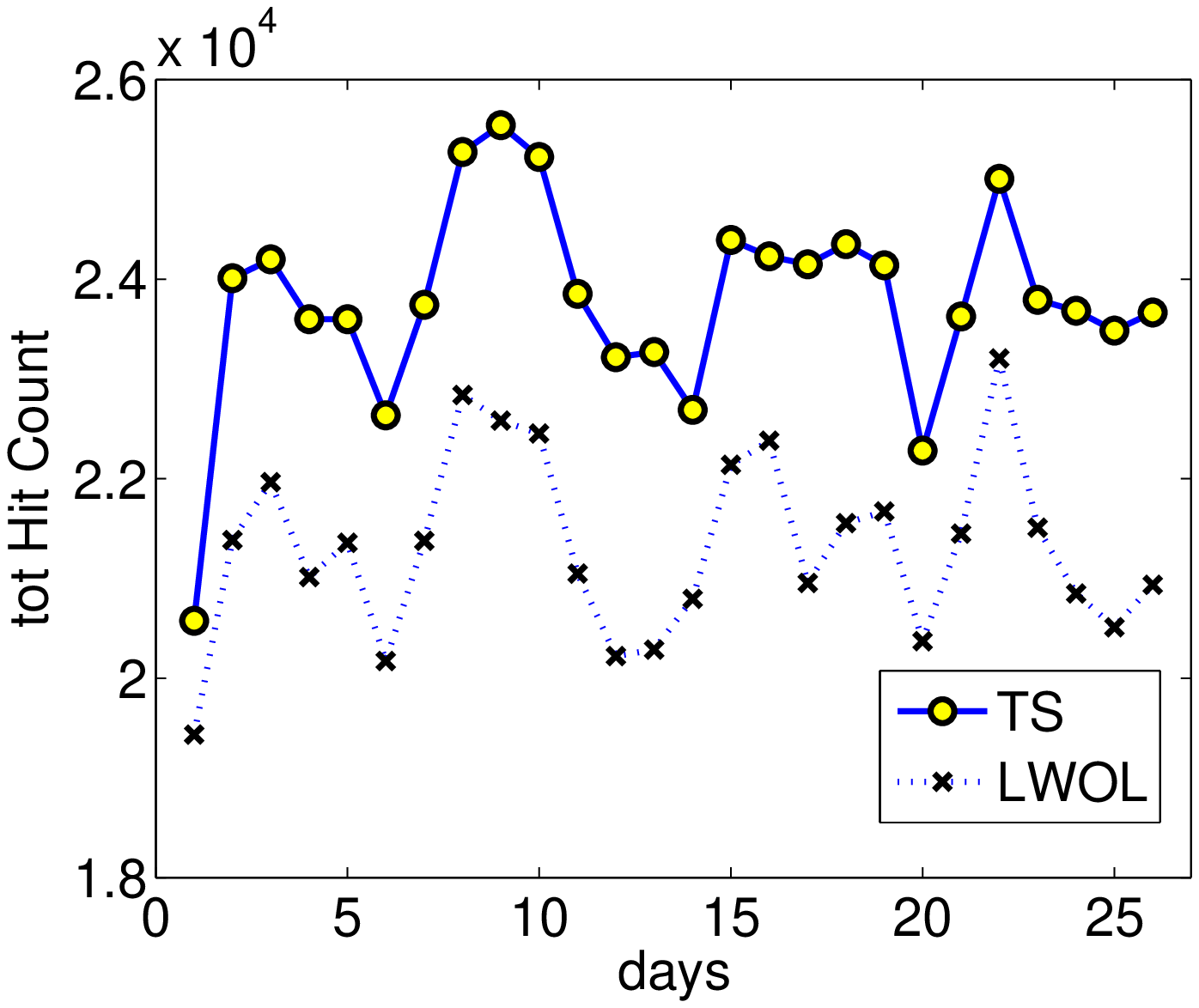}}
\subfigure[Global (neighborhood) approaches : $K$NN (``on TS'',``on Train''), EWMA-CA, GWOL]{
\includegraphics[width=58mm]{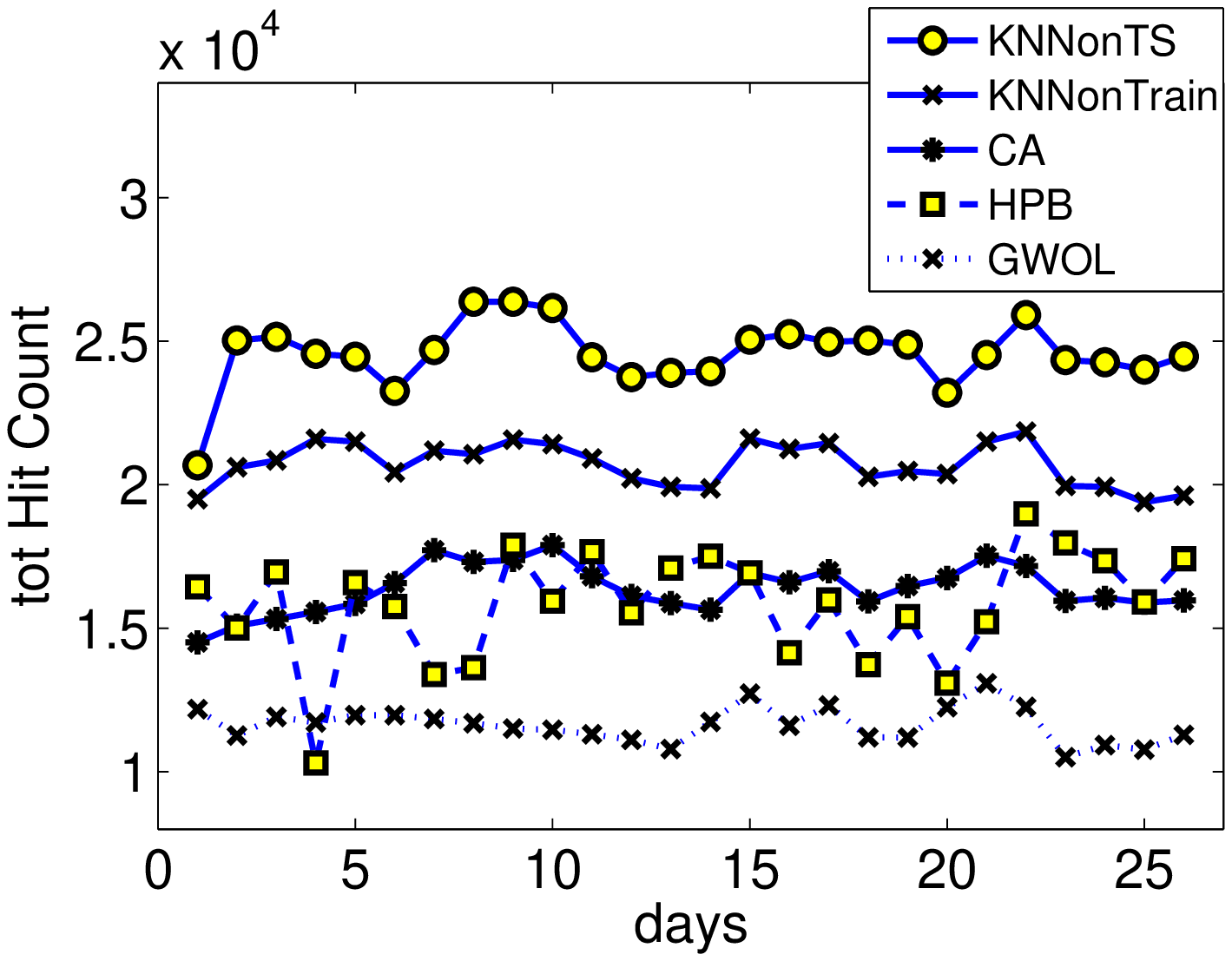}}
\subfigure[Proposed combined method (TS+$K$NN+CA) vs. state-of-the-art (HPB) and baseline (GWOL)]{
\includegraphics[width=58mm]{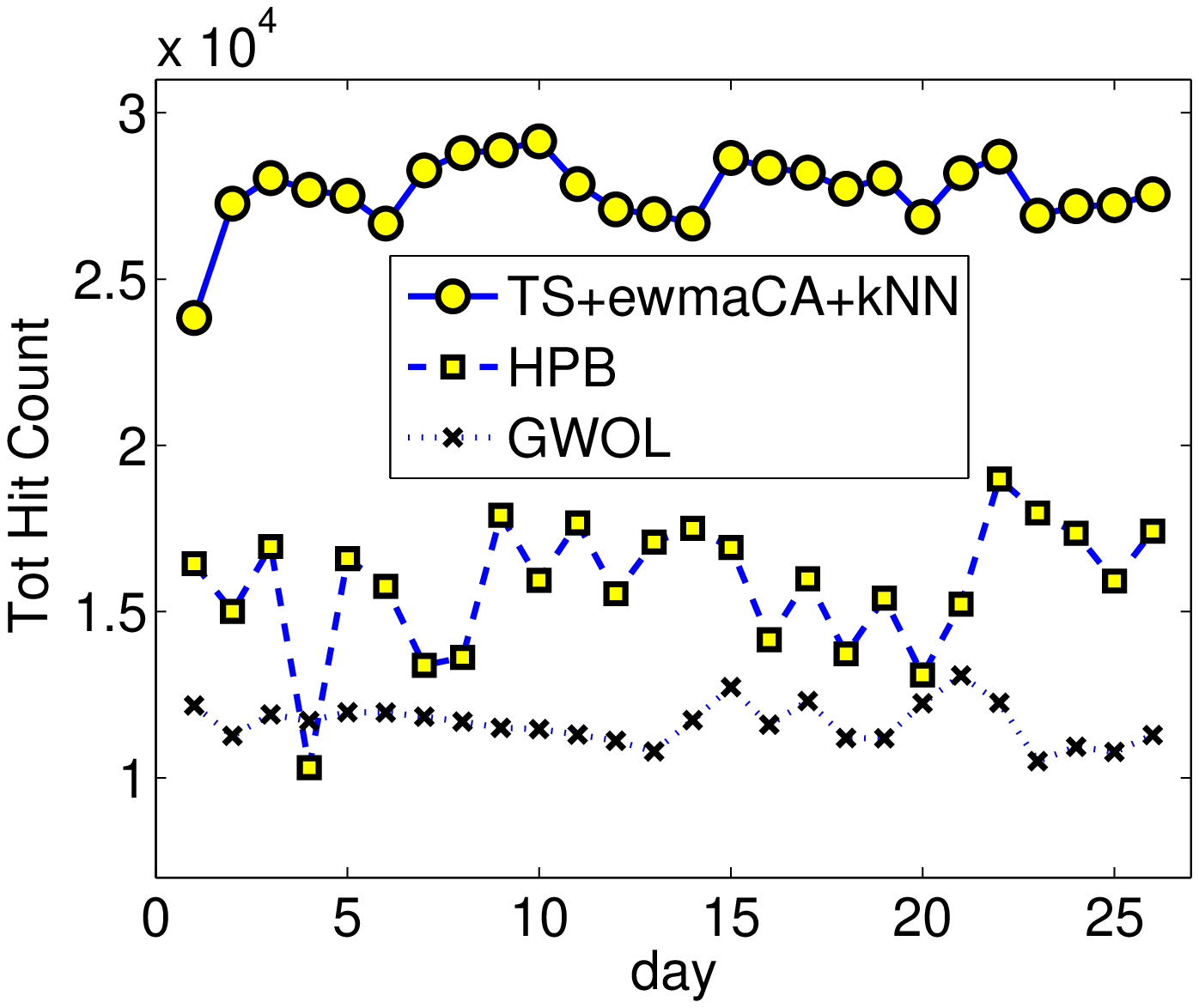}}
\caption{Evaluating the performance (total hit count) of different individual methods, our proposed combined method (TS+KNN+CA) and baselines methods.}
\label{fig:comparison}
\end{figure*}

In Fig. \ref{fig:comparison}(a) we plot and compare the total hit count of {\em local schemes}, namely TS and LWOL. Their performance oscillates based on the specific (training and testing) data available on different days. However, we can see that the TS approach consistently outperforms LWOL over all days. This is expected since  the TS includes LWOL as a special case  when the training window is equal to a single time slot.

In Fig. \ref{fig:comparison}(b), we compare the hit count of {\em global schemes} that use information from different networks, namely GWOL, HPB, EWMA-CA, and $k$NN.
We implemented the relevance propagation used in HPB with parameter 0.5.
As noted in \cite{Zhang2008}, the average improvement of HPB over GWOL is $\sim$36\%. The EWMA-CA algorithm has on average the same performance as HPB. However, (i) its performance is more consistent across time than HPB and (ii) the two methods capture different concepts of neighborhood (victim neighborhood in HPB vs. joint victim-attacker neighborhood in EWMA-CA). Thus, they potentially capture different set of attackers, which explains the difference in performance. Finally, we plot both prediction models for $k$NN: ``$k$NN on Train'' in Eq. (\ref{knn}) (where $k$NN is run on top of the last day's logs), ``$k$NN on TS'' in Eq. (\ref{knnTS}) (where $k$NN is run on top of the TS predictions). We set $k= 25$. $k$NN schemes outperforms other neighborhood schemes, mainly thanks to the notion of similarity that accounts for simultaneous attacks. 
Computing $k$NN on top of the TS prediction results in further improvement.

In Fig.\ref{fig:comparison}(c), we show the total hit count achieved by {\em our proposed combined scheme of Eq. (\ref{combination}}), which blends together TS , $k$NN (on TS) and (EWMA) CA and we compare it to the state-of-the-art method (HPB). This figure shows essentially the {\em main result} of this paper. Our scheme outperforms HPB significantly (up to 70\% over the hit count of HPB with an average improvement over 57\%) and consistently (in every day of October 2008). We also show the more traditional baseline GWOL that performs even worse than HPB.

We also investigated the reasons behind this improvement in more detail.
  First, we looked at the set (not only the number) of attackers predicted by each individual method. Each method provides every contributor with a customized blacklist that successfully predicts some attackers. Besides predicting a common set of attackers, the three different methods successfully predict disjoint sets of attacks of significant size. {\em E.g.,}
  TS and EWMA-CA successfully predict a common set of 9.9 K attackers; however, EWMA-CA alone captures an additional 6.1 K attackers that the TS alone cannot. This motivates the combination of these three prediction schemes so that they can complement each other and explains the hit count improvement when combining them.
Second, adding new schemes in the combination improves the hit count but has diminishing returns, as it is also the case in traditional recommendation systems \cite{netflix, ensemble}. In particular, adding EWMA-CA to TS results in a 12\% average hit count improvement; adding $k$NN to the combination TS + EWMA-CA results in only 6\% average improvement.
This suggests that incorporating  additional neighborhood schemes into the equation would likely give modest improvement.

\begin{figure}[t!]
\begin{center}
\includegraphics[scale=0.40]{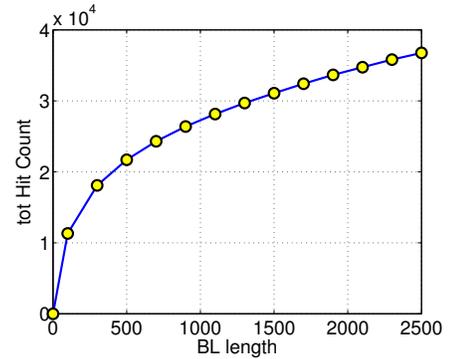}
\caption{ Hit count as function of the blacklist length}
\label{fig:BLsize}
\end{center}
\end{figure}

In Fig. \ref{fig:BLsize} we analyze the performance of our algorithm as a function of the blacklist length. As we expect both the hit count and the prediction rate increases with the blacklist length. The larger relative increase occurs with a blacklist of length 3--500. In fact, a blacklist of length 500 has on average a prediction rate of about 50\%. While a blacklist 5 times longer, corresponding to 2500 entries, has a prediction rate of about 59\%. This suggests that the best length for our predictive blacklist is about 500 entries.

\subsection{\label{sec:windows}Training and Testing Windows}

\begin{figure}[t!]
\centering
\subfigure[Hit count vs. training window length]{ 
\includegraphics[width=55mm]{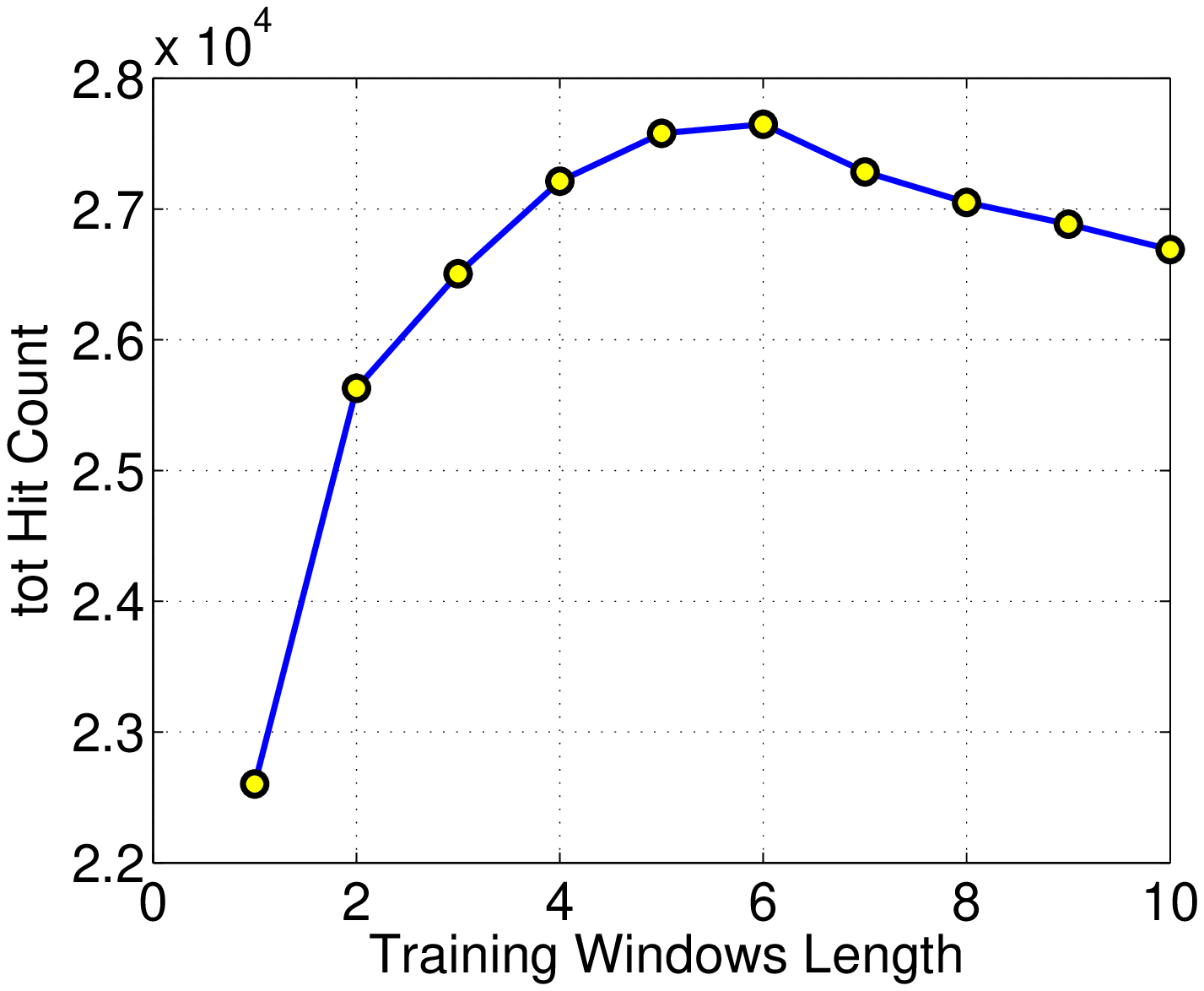}}
\subfigure[Hit count vs. test window length]{
\includegraphics[width=55mm]{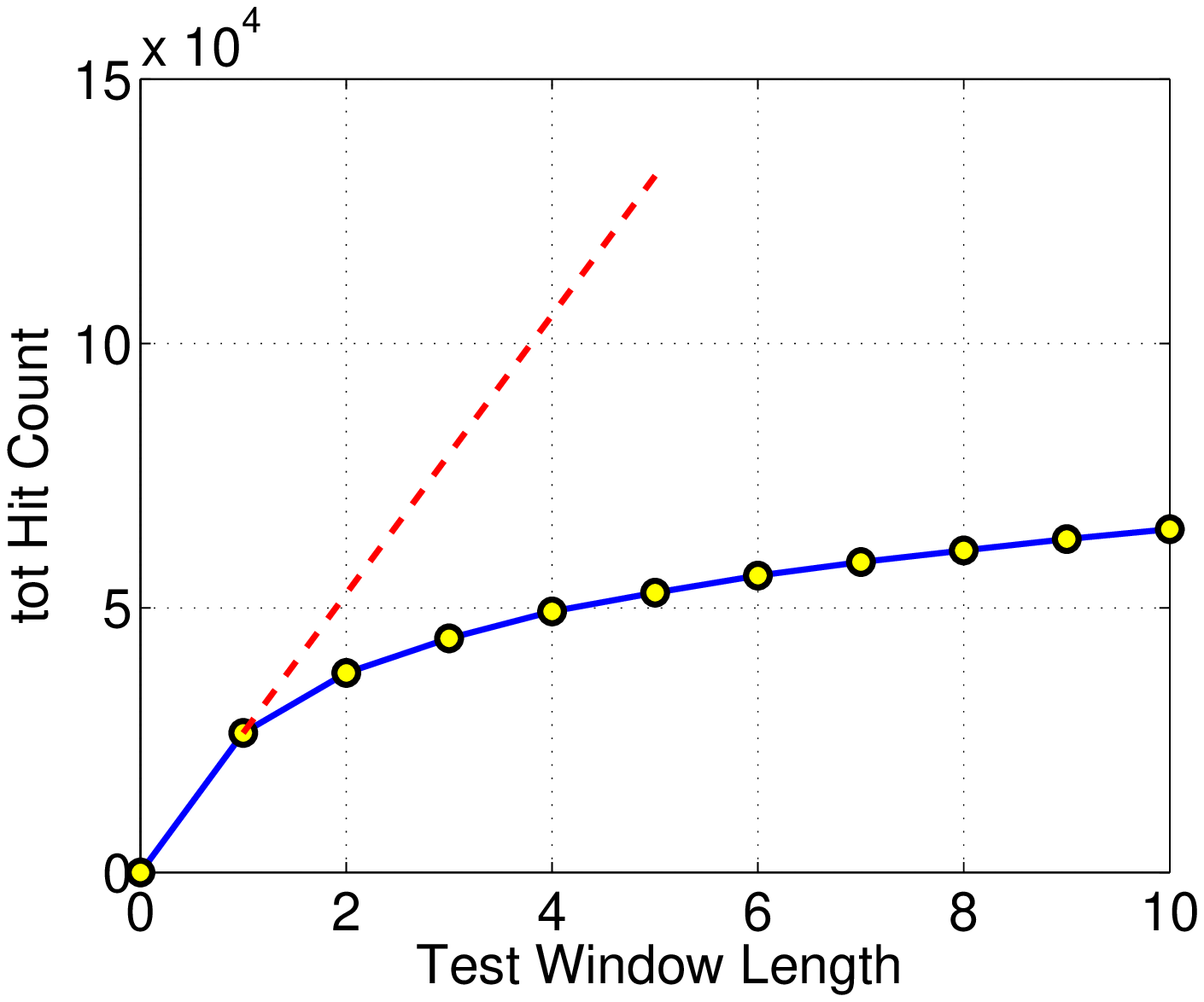}}
\subfigure[Prediction rate vs. test window length]{
\includegraphics[width=57mm]{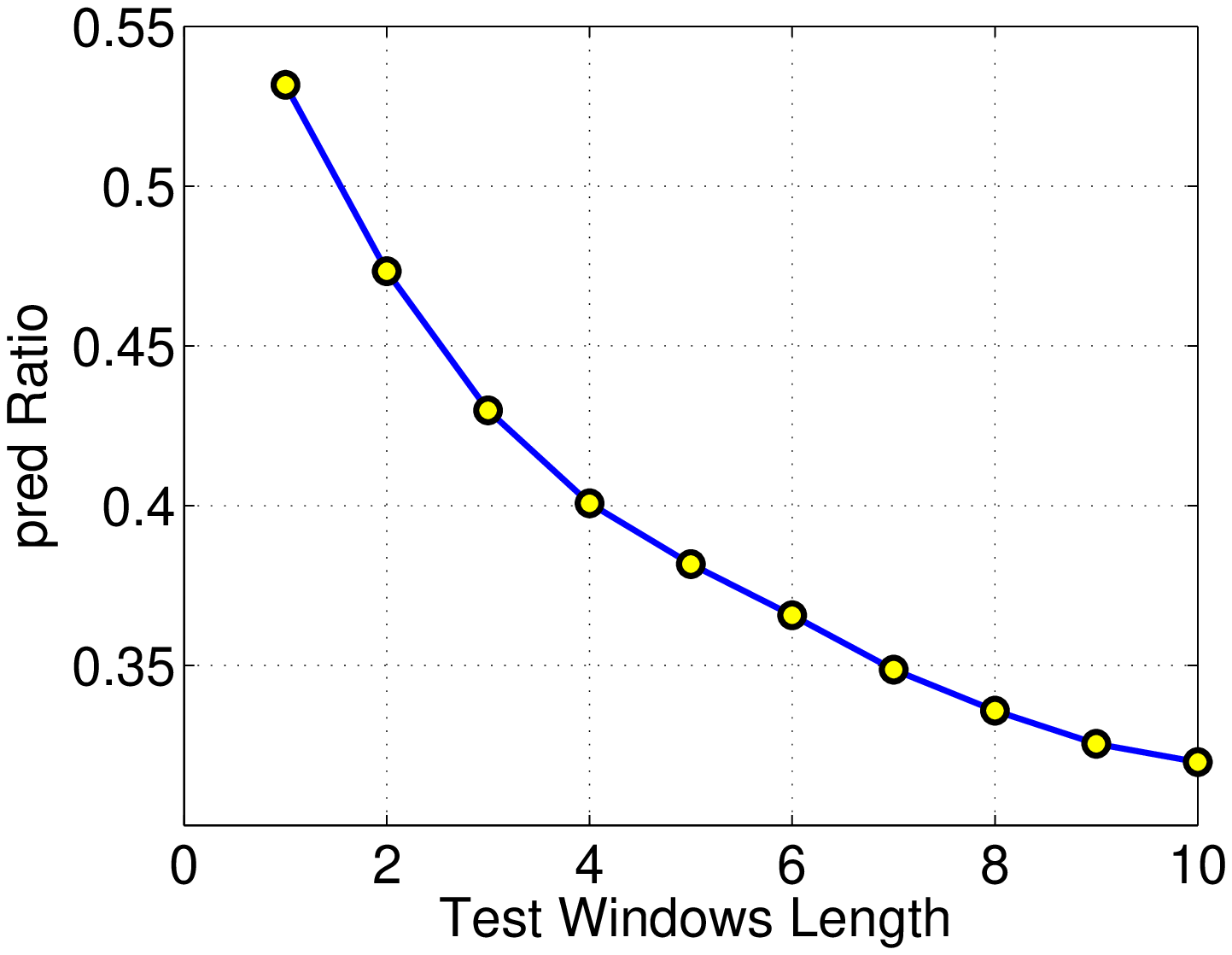}}
\caption{Tuning the training and testing window of our proposed combined method based on our data. Every point on these plots represents the average total hit count over 7 consecutive days. At the end, we chose a training window of 5 days and a test window of 1 day.}
\label{fig:windows}
\end{figure}

%



%

Throughout the paper we used training and testing windows of 5 and 1 days respectively.
Fig. \ref{fig:windows}  shows the performance of our prediction scheme as a function of the length of these windows
and justifies these choices.

We observe that when the training window is too short, the benefit of the time series model is limited by the few available observations.
When the  training window is too long, it introduces correlation between remote past and recent activities, which was not the case in our data analysis ({\em e.g.,} Fig. \ref{fig:data3}(b)).
Fig.\ref{fig:windows}(a) clearly shows this trade-off. The performance of our prediction algorithm first increases with the training windows then it decreases when the windows is more than 6-day long. In fact, the curve empirically shows that our scheme achieves the optimal performance when trains on 5-6-day data. 

In Fig. \ref{fig:windows}(b), we plot the hit count as a function of the length of the testing window. Here, we make two main observations: (1) by increasing the testing window from 1 to 10, the hit count is more than doubled; and (2) this improvement, although quite significant at first, is much smaller than the hit count we would have by running the prediction from scratch every day (dashed line). We also looked at the ratio of the hit count over the upper bound for prediction (omitted for lack of space) and we found that this relative performance metric decreases with the testing window. This indicates that a short testing window is preferable, or in other words, prediction should be trained/refreshed often.



\subsection{Robustness against Pollution of the Training Data}

%

  Large-scale repositories that collect firewall and IDS logs from different networks (contributors), such as {\tt Dshield}, are naturally prone to include a certain number of false alerts, as the repository has no control over the contributed logs. False alerts may be either due to errors in the configuration of the IDS of a contributor (pollution) or due to a malicious contributor trying to mislead our prediction (poisoning). It turns out that using a combination of diverse prediction methods increases the robustness against both problems.

\begin{figure}
\begin{center}
\includegraphics[width=51mm]{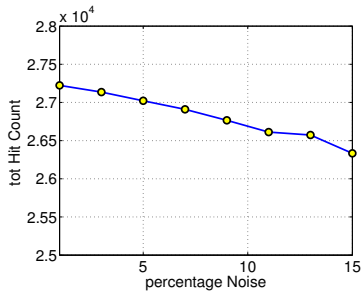}
\caption{\label{fig:percNoise} Robustness of the combined method in the presence of pollution of the training data. The total hit count decreases much slower than the random noise (\% of the total number of reports). }
\end{center}
\end{figure}

  {\bf Pollution.} To quantify how random false positives affect the prediction accuracy of our combined method,
we artificially generated fake reports, which are distributed over all contributing networks proportionally to the number of real reports they submitted. We vary the amount of total fake reports generated (noise) from 1\% to 15\%.  Fig.\ref{fig:percNoise} shows the results. We observe that the hit count decreases slower than the pollution rate, {\em e.g.,} by less than 4\% when the pollution rate is 15\%. This can be explained as follows. False alerts generated at different networks are unlikely to affect neighborhood models because they usually correspond to different sources reported by different contributing networks, which does not introduce correlation between victims. in order to introduce such correlation, fake reports should have not only the same source but also a similar time stamp to affect the $k$NN model presented. Finally, if a source is falsely reported over several days by the same victim, this can affect only the blacklist customized for that specific victim, since the time series prediction is specifically computed for each victim network.


{\bf Poisoning.} Evading our combined prediction  is difficult for an attacker and comes at the cost of limiting the attack impact. Indeed, an attacker must avoid both the time series prediction and the two neighborhood-based methods. To mislead the time series, an attacker can limit traffic towards the target network. In fact, even activities that have low intensity but are persistent over time will be revealed by the time series model. 
 Instead, an attacker might attack different networks for a short time, a behavior that will be captured by the neighborhood-based models, which focus precisely on this type of behaviors.

%% file: conclusion.tex
\section{Summary and Future Work}
%

In this paper, we studied the problem of predicting future malicious activity (through ``predictive blacklists'')  given past observations
(available through a shared repository of logs from different victims/contributors).
We framed the problem as an implicit recommendation system, which paves the way to the application
of powerful machine learning methods. Within this framework, we 
also proposed a specific prediction method, which is a linear blend of three different algorithms: 
a time series model to account for the temporal dynamics and two neighborhood-based models. 
The first neighborhood model, is an adaptation of $k$NN model for attack prediction and focuses on capturing similarity between victims being attacked by the same sources, preferably at the same time. 
The second is a co-clustering algorithm that automatically discovers a group of attackers that 
 attack a group of victims at the same time. 

We analyzed a real dataset of 1-month logs from {\tt Dshield.rg}, consisting of 
of 100s of millions network security logs contributed by 100s of different networks.
We evaluated our proposed algorithms over this dataset and showed significant improvement over the state-of-the-art attack prediction methods.
Our combined method improves significantly not only the prediction accuracy 
but also the robustness against pollution/poisoning of the dataset.


Despite our performance improvement and methodological development over the state-of-the-art, we believe that this work only scratches the surface of the complicated attack prediction problem. Our analysis shows that even larger improvements can be obtained ({\em i.e.,} there is still a gap between our method and the upper bound). There are several directions for future work: (a) incorporate the effect of other fields/dimensions of the dataset (such as destination port ID) into our prediction  model;
(b) add new algorithms in our combination that capture different effects ({\em e.g.,} latent factor models
could capture global behavior); (c) build a prototype. 


